\documentclass[a4paper]{llncs}

\usepackage{graphicx}
\usepackage{vdmlisting}
\usepackage{listing}
\usepackage{hyperref}
\usepackage{url}
\lstdefinestyle{overtureLanguageStyle}{basicstyle=\footnotesize\ttfamily,
			numbers=left,
                        frame=trBL, 
                        tabsize=2, 
                        linewidth=\textwidth,
                        showstringspaces=false, 
                        captionpos=b,
                        frameround=fttt, 
                        aboveskip=2mm,
                        belowskip=2mm,
                        framexleftmargin=0mm, 
                        framexrightmargin=0mm,
                        escapeinside={(*@}{@*)},
                        language=VDM_SL}
\lstdefinelanguage{python}{ 
  backgroundcolor={\color[gray]{1}},
  basicstyle=\small\ttfamily,
  sensitive=true, 
  morestring=[s]{"}{"}, 
} 
\lstdefinelanguage{vdmsl}{ 
  backgroundcolor={\color[gray]{1}},
  basicstyle=\small\ttfamily,
  sensitive=true, 
  morestring=[s]{"}{"}, 
} 

\title{Specification Slicing for VDM-SL}

\titlerunning{22nd Overture Workshop, 2024}

\author{
Tomohiro Oda\inst{1} \and 
Han-Myung Chang\inst{2}
}

\authorrunning{Oda, T.; Chang, H.M.}

\institute{Software Research Associates, Inc.~(\email{tomohiro@sra.co.jp})
\and Nanzan University~(\email{chang@nanzan-u.ac.jp}) 
}
\begin{document}

\maketitle

\begin{abstract}
The executable specification is one of the powerful tools in lightweight formal software development. 
VDM-SL allows the explicit and executable definition of operations that reference and update internal state through imperative statements. 
While the extensive executable subset of VDM-SL enables validation and testing in the specification phase, it also brings difficulties in reading and debugging as in imperative programming.
In this paper, we define specification slicing for VDM-SL based on program slicing, a technique used for debugging and maintaining program source code in implementation languages.
We then present and discuss its applications.
The slicer for VDM-SL is implemented on ViennaTalk and can be used on browsers and debuggers describing the VDM-SL specification.
\end{abstract}

\section{Introduction}
VDM-SL~\cite{VDM10LRM} is a formal specification language with an extensive executable subset.
A system is modelled as a collection of inter-connected modules, each of which has an internal state and operations that reference or update the internal state.
Internal states and operations are defined within modules using mathematical language elements such as types, constants, and functions.
Operations can be defined implicitly and abstractly by preconditions and postconditions, or explicitly and executably by imperative statements.

By executing the specification, testing techniques such as unit testing and combinatorial testing can be applied~\cite{CombinatorialTesting,VDMUnit}.
It is also possible to validate whether the VDM-SL specification conforms to actual usage scenarios using interpreters~\cite{WiderRangeOfStakeholders} or code generators~\cite{CodeGenerator}.
Along with the benefits of testing and validation, defining operations using imperative statements brings the complexity of imperative programming to the specification phase.
Because the execution of each statement is affected by preceding destructive assignments to variables, control structures, and calls to other operations, it is less obvious as to which part of the effect of an operation a statement contributes and in what way.

Program slicing~\cite{ProgramSlicing} is a technique to extract relevant parts of a program source based on the given slicing criterion.
There are variations of program slicing techniques, and among them, we refer to static backward slicing which statically analyzes the given program source and extracts parts of the program that may affect the value of the specified variable or expression.
Static backward slicing is used for debugging by narrowing the possible causes of the issue down to a slice.
Slicing is also helpful in understanding imperative programs that manipulate multiple variables by separating the program into slices for each variable.

Oda and Araki proposed specification slicing, an application of the slicing technique to Z notation, a formal specification language based on the set theory and first-order logic~\cite{ZSlicing}.
Although Z notation has a limited executable subset, it does not have imperative statements but declarations of variables and constraints on them.
Slicing Z specification relies on transitive tracing of constraints on variables and it is hard to identify a minimal subset of constraints that affect possible values of a variable.
On the other hand, VDM-SL has imperative statements, to which general program slicing techniques can be directly applied.

In Section \ref{sec:definition-and-extraction}, we define slice extraction algorithm for VDM-SL.
We introduce expected applications in Section \ref{sec:applications}, and then discuss the implementation of a slicer and tools that use the slicer in ViennaTalk.
Section \ref{sec:concluding-remarks} concludes and explain future work.

\section{Definition and Extraction}
\label{sec:definition-and-extraction}

\begin{figure}
\begin{center}
\includegraphics[width=0.5\textwidth]{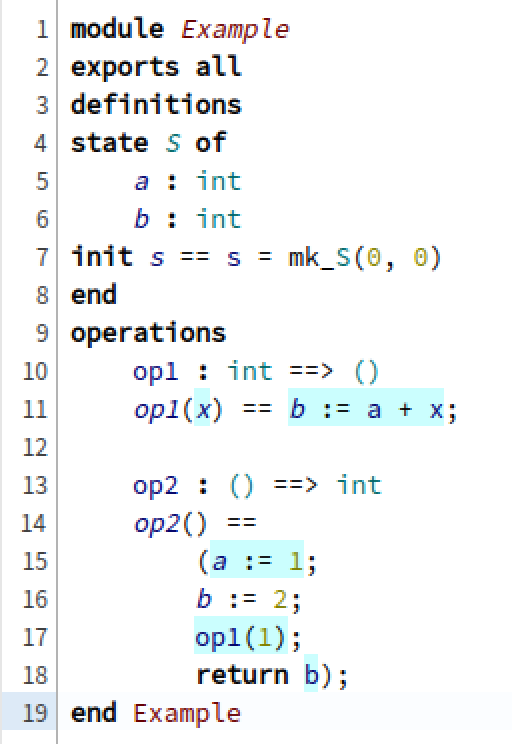}
\end{center}
\caption{An example static backward slice}
\label{fig:example-slice}
\end{figure}

In this paper, we define a static backward slice, or simply a {\it slice} in this paper, of a specification {\tt s} for a slicing criterion {\tt C = (o, e)}.
A slice is a subset of the AST (Abstract Syntax Tree) nodes of the specification {\tt s} that may influence the value of the expression {\tt e} in the execution of the operation {\tt o}.
In Fig. \ref{fig:example-slice}, the slice for the return value {\tt b} in the execution of the operation {\tt op2} is highlighted in cyan.

In Fig \ref{fig:example-slice}, the operation {\tt op2} assigns the value {\tt 2} to the state variable {\tt b} (Line 16), but later the variable {\tt b} is overwritten in the operation call to {\tt op1} (Line 11).
As a result, the value of the variable {\tt b} in Line 18 does not depend on the assignment statement in Line 16 and therefore Line 16 is not included in the slice.
Instead, Line 11 assigns the value of {\tt a + x} to the variable {\tt b}, which puts the assignment statement into the slice.
The state variable {\tt a} referenced in Line 11 is assigned in Line 15, which is also included in the slice.
In this section, the overview of the slice extraction method will be explained.

\subsection{Slicer}
In general, backward program slicing identifies dependency among the AST nodes, and transitively follows the dependency from the slicing criterion.
Our slicing algorithm uses the variables listed in Fig. \ref{fig:slicer-state} to identify and follow the dependency.
The variables {\tt criteria} and {\tt toplevel} are set from the slicing criterion.
The variable {\tt toplevel} is typed as an option type just for initialization.
The variable {\tt agenda} holds the dependency that the slicer is now following.
At the initialization of the slicer, the variable {\tt agenda} is copied from the variable {\tt criteria}.
The variables {\tt reads} and {\tt writes} store read and write accesses in the current statement, and the variable {\tt slice} stores the AST nodes found in the slice so far.
The variables {\tt read}, {\tt writes} and {\tt slice} are initialized to empty.
\begin{figure}
\begin{vdmsl}
state Slicer of
  criteria : set of AST
  toplevel : [AST]
  agenda : set of AST
  reads : set of AST
  writes : set of AST
  slice : set of AST
init s == s = mk_Slicer({}, nil, {}, {}, {}, {})
end
operations
setCriterion: AST * set of AST ==> ()
setCriterion(o, e) == 
  (criteria := e; 
  toplevel := o; 
  agenda := e; 
  reads := {}; 
  writes := {}; 
  slice := {});
\end{vdmsl}
\caption{Pseudo-code of the variables in the slicer and their initialization}
\label{fig:slicer-state}
\end{figure}

The slicer processes the AST of the operation definition starting at the slicing criterion and in the reverse order of interpretation.
At each AST statement node, the slicer executes the pseudo-code shown in Fig. \ref{fig:slicer-overview}.
The slicer updates  {\tt reads} and {\tt writes} depending on each AST node (the cases statement in the Fig \ref{fig:slicer-overview}).

The operation {\tt process\_dependency} is a common procedure to check whether or not any AST node in {\tt agenda} is also in {\tt writes}, which means that the slicer identified a data dependency.
If so, the slicer updates {\tt agenda} so that the slicer retracts interests in the written nodes and the nodes in {\tt reads} will be followed.
The AST node is influential to the slicing criterion and therefore added to {\tt slice}.
In the following sections, specific processes on major kinds of AST nodes will be explained.

\begin{figure}
\begin{vdmsl}
operations
process: AST ==> ()
process(node) == 
  (cases node:
    /* AST node specific processes */
  end;

process_dependency: AST ==> ()
process_dependency(node) ==
  (let common = agenda inter writes in
    if common <> {}
    then 
        (agenda := (agenda \ common) union read;
         slice := slice union {node});
  writes := {};
  reads := {});
\end{vdmsl}
\caption{Pseudo-code of the overview of how the slicer process each AST node}
\label{fig:slicer-overview}
\end{figure}

\subsection{Pure expressions}
The most expressions in VDM-SL do not change the state but may refer to state variables.
An interpreter evaluates an expression by evaluating subexpressions and then computes the value of the entire expression using the values of the subexpressions.
A slicer works in the other direction.
Fig. \ref{fig:slicer-expression} shows a pseudo-code for binary operator expressions to be filled in as a case-alternative into the cases statement in Fig \ref{fig:slicer-overview}.
The slicer adds the entire expression node into the {\tt writes} variable to mean that the value of the entire expression will be computed and will be passed to the caller's context.
The slicer also adds the subexpressions into the {\tt reads} variable.
By calling {\tt process\_dependency}, the slicer identifies dependency and updates the {\tt slice} and {\tt agenda} for further slicing operations.
If the expression node is in the {\tt agenda} variable, the slicer then identifies the data dependency from the expression node to the subexpression.
The subexpression will be added into the {\tt agenda} variable.
The slicer then processes subexpressions in order, and on each subexpression, the slicer will add the subexpression into the {\tt write} variable.
This recursion chains dependency from the expression to its subexpressions.

The recursion into subexpressions will terminate at atomic expressions such as name references and literal values.
The name nodes and literal nodes under an expression are added in the caller's context, and the slicer will do nothing to process name references and literal values.

\begin{figure}
\begin{vdmsl}
    mk_BinExp(op, exp1, exp2) -> 
      (writes := writes union {node}; 
      reads := reads union {exp1, exp2}; 
      process_dependency(node);
      process(exp2); 
      process(exp1)),
    mk_Literal(v) -> skip,
    mk_Name(identifier) -> skip,
\end{vdmsl}
\caption{Pseudo-code of processing binary operator expressions}
\label{fig:slicer-expression}
\end{figure}

\subsection{Assignments and pattern matchings}
Assignments and pattern matchings are major sources of data dependency.
Fig \ref{fig:slicer-assign} shows how a slicer finds read and write accesses in an assignment statement.
In VDM-SL, the left-hand side of an assignment statement is not only a simple variable name but a state designator which may involve field references and map/sequence references.
The slicer uses two utility functions {\tt var} and {arg\_expressions} to extract the variable and map/sequence reference arguments.
The slicer adds the variable node in the state designator {\tt var(state\_designator)} to the {\tt writes} variable, and also adds all arguments in the state designator ({\tt args}) to the {\tt reads} variable along with the right-hand side expression.
The slicer then identifies and follows the dependency.
\begin{figure}
\begin{vdmsl}
    mk_Assign(state_designator, expression) -> 
      let args = elems arg_expressions(state_designator) in
        (writes := writes union {var(state_designator)};
        reads := reads union {expression} union args;
        process_dependency(node);
        for arg in reverse args do process(arg);
        process(expression)),
\end{vdmsl}
\caption{Pseudo-code of processing assignment statements}
\label{fig:slicer-assign}
\end{figure}

Pattern matching is a prominent language feature of VDM-SL.
While identifiers are presented as the left-hand side of local definitions and formal parameters of functions and/or procedures in many programming languages, VDM-SL allows rich patterns, such as set union patterns, record patterns and match value patterns, to be placed as the left-hand side of local definitions and formal parameters.
Identifiers in those patterns should also be added to the {\tt writes} variable.
Each match value pattern, denoted by an expression enclosed by a pair of parentheses,  matches to the identical value.
The expressions in match value patterns influence which value the other pattern identifiers match, and therefore should be added to the {\tt reads} variable.
Fig \ref{fig:slicer-pattern} shows the process of pattern identifier, set union patterns and match value patterns.

\begin{figure}
\begin{vdmsl}
    mk_PatternIdentifier(identifier) -> 
      (writes := writes union {node};
      process_dependency(node)),
    mk_SetUnionPattern(pattern1, pattern2) -> 
      (writes := writes union {node};
      reads := reads union {pattern1, pattern2};
      process_dependency(node);
      process(pattern2);
      process(pattern1)),
    mk_MatchValuePattern(expression) -> 
      (writes := writes union {node};
      reads := reads union {expression};
      process_dependency(node);
      process(expression)),
\end{vdmsl}
\caption{Pseudo-code of processing patterns}
\label{fig:slicer-pattern}
\end{figure}

\subsection{Sequential executions, branches, loops and calls}
VDM-SL has statements for control structures.
Fig. \ref{fig:slicer-sequential-branch} shows pseudo-code for the block statement and the if statement.
A block that consists of two or more statements optionally with local variable declarations can be processed simply by processing its substatements in the reverse order.
For a branch, the slicer processes all possible execution paths in a conditional statement and merges them by taking unions of {\tt agenda}, {\tt reads} and {\tt writes} from the execution paths.

\begin{figure}
\begin{vdmsl}
    mk_Block(dcl_statement, statements) -> 
      (for statement in reverse statements do process(statement);
      if dcl_statement <> nil then process(dcl_statement)),
      process_dependency(node)),
    mk_BinIfStatement(expr,then_statement, else_statement) -> 
      (dcl saved_agenda : set of AST := agenda, 
        saved_reads : set of AST := reads,
        saved_writes : set of AST := writes,
        branch_agenda : set of AST, 
        branch_reads : set of AST,
        branch_writes : set of AST;
      if else_statement <> nil then process(else_statement);
      if agenda <> saved_agenda then 
        agenda := agenda union {expr};
      process(expr);
      branch_agenda := agenda;
      branch_reads := reads;
      branch_writes := writes;
      agenda := saved_agenda;
      reads := saved_reads;
      writes := saved_writes);
      process(then_statement);
      if agenda <> saved_agenda then 
        agenda := agenda union {expr};
      process(expression);
      agenda := agenda union branch_agenda;
      reads := reads union branch_reads;
      writes := writes union branch_writes),
\end{vdmsl}
\caption{Pseudo-code of processing sequential execution and branch}
\label{fig:slicer-sequential-branch}
\end{figure}

If statements in VDM-SL may have {\tt elseif} clauses, which allows a chain of multiple conditionals.
The pseudo-code shown in Fig. \ref{fig:slicer-sequential-branch} shows how to process a simple if-then-else statement without elseif clauses because if statements with elseif clauses can be replaced with their equivalent if-then-else statement.
In the pseudo-code, the slicer processes two paths: {\tt expr} $\rightarrow$  {\tt then\_statement} and {\tt expr} $\rightarrow $ optional {\tt else\_statement}, and then they are merged by taking unions.
A control dependency is identified when either statement in the then or else clause changes the {\tt agenda} variable.
The basic idea is, that if there is a slice in one or more branches, the condition expression will also be put into the slice.

Fig. \ref{fig:slicer-loop} shows a pseudo-code for slicing while statements.
The slicer first processes the conditional expression because the interpreter terminates the while loop with the conditional expression being false.
The slicer repeats processing the loop content and the conditional until the result converges.
The control dependency is identified if the {\tt agenda} variable changes by processing the loop content.
The resulting {\tt agenda}, {\tt reads} and {\tt writes} are union of all iterations.

\begin{figure}
\begin{vdmsl}
    mk_WhileStatement(expr, statement) -> 
      (dcl saved_agenda : set of AST, saved_reads : set of AST,
        saved_writes : set of AST, branch_agenda : set of AST, 
        branch_reads : set of AST, branch_writes : set of AST;
      if else_statement <> nil then process(else_statement);
      if agenda <> saved_agenda then 
        agenda := agenda union {expr};
      process(expr);
      saved_agenda := agenda; 
      saved_reads := reads; saved_writes := writes;
      branch_agenda := agenda;
      branch_reads := reads; branch_writes := writes;
      process(statement);
      if agenda <> saved_agenda then
        agenda := agenda union {expr};
      process(expr)
      while agenda psubset branch_agenda or 
            reads psubset branch_reads or writes <> branch_writes 
      do
        (branch_agenda := branch_agenda union agenda;
        branch_reads := branch_reads union reads;
        branch_writes := branch_writes union writes;
        saved_agenda := agenda;
        saved_reads := reads;
        saved_writes := writes;
        process(statement);
        if branch_agenda <> saved_agenda then
          agenda := agenda union {expr};
        process(expr));
      agenda := branch_agenda;
      reads := branch_reads;
      writes := branch_writes),
\end{vdmsl}
\caption{Pseudo-code of processing while loops}
\label{fig:slicer-loop}
\end{figure}

In VDM-SL, an apply expression may cause a change of state when the callee is an operation.
A function also may refer to a value definition which may influence the slicing criterion.
A slicer traverses function definitions and operation definitions when the slicer encounters an apply expression or an operation call.


\section{Applications}
\label{sec:applications}

Program slicing tools have been used in debugging and maintenance to narrow the scope of the source in concern.
In this section, we describe how specification slicing can help debugging and refactoring tasks.

\subsection{Debugging}

While program code in programming languages is oriented to execution, executable specifications aim at defining important properties and organizations of functionalities.
Even though explicit operations with statements can run without preconditions and postconditions, preconditions and postconditions are the main concerns of specification engineers.
In this section, we start with an erroneous specification that causes a postcondition violation and find the bug using specification slicing.

\begin{figure}
\begin{vdmsl}
state MemberBook of
    EmailBook : map Id to Email
    NameBook : map Id to Name
    NextId : Id
inv mk_MemberBook(emails, names, next) ==
    next not in set dom emails and next not in set dom names
init s == s = mk_MemberBook({|->}, {|->}, 1)
end
operations
    register : Name * [Email] ==> Id
    register(name, email) ==
        (dcl i:Id := NextId;
        NextId := NextId + 1;
        NameBook(i) := name;
        if
            email <> nil
        then
            (i := NextId;
            NextId := NextId + 1;
            EmailBook(i) := email);
        return i)
    post 
        NameBook = NameBook~ munion {RESULT |-> name}
        and (email = nil and EmailBook = EmailBook~
            or email <> nil and 
              EmailBook = EmailBook~ munion {RESULT |-> email});
\end{vdmsl}
\caption{Erroneous specification of MemberBook}
\label{fig:memberbook-erroneous}
\end{figure}

Fig. \ref{fig:memberbook-erroneous} shows a part of an erroneous specification of a member management system.
The system issues an ID for each member and records the name and optionally an email address associated with the ID.
To generate unique IDs, the system holds the {\tt NextId} state variable, and the invariant (Line 5-6) states that the value of the {\tt NextId} must not be already associated with any name or email address.

The {\tt register} operation accepts two arguments, {\tt name} and {\tt email}.
The operation picks the {\tt NextId} and increments it to keep it unique (Line 12-13), and associates the name to the id (Line 14).
In VDM-SL, Line 14 does not overwrite the existing map value but assigns a new map value {\tt NameBook munion \{i |-> name\}}.
The operation then processes the email in the same manner if the argument {\tt email} is not nil.
Finally, the operation returns the value of {\tt id} that the given name and optional email address are associated with, as postcondition (Line 22-26) asserts.

\begin{figure}
\begin{center}
\includegraphics[width=1\textwidth]{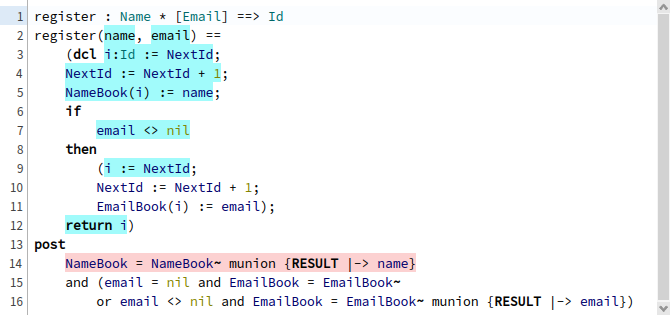}
\end{center}
Highlighted in red: the violated assertion.\\
Highlighted in cyan: the slice for the violated assertion.
\caption{Slice for  the violated postcondition}
\label{fig:memberbook-debug-slice}
\end{figure}

When {\tt register("John Doe", "jd@example.com")} is evaluated as a test, a postcondition violation occurs.
The statements in the {\tt register} operation do not work as intended.
The violated postcondition is shown in red in Fig. \ref{fig:memberbook-debug-slice}.

To spot the erroneous statements, slice for the violated postcondition {\tt  NameBook = NameBook\~{} munion \{RESULT $\mapsto$ name\}}  (shown in cyan in Fig \ref{fig:memberbook-debug-slice}) should contain the erroneous statements.
The slice should include the parts that contribute to how the name is stored.
The slice looks suspicious because the condition expression of the if statement (Line 16) is included.
Why the email should be involved in storing the name?
In Line 9, the assignment to the local variable {\tt i} is also in the slice.
Yes, Line 9 modifies the return value, and therefore should be in the slice.
Line 9 is erroneous because it makes the name no longer associated with the returned ID.
Fig. \ref{fig:memberbook-corrected} shows the corrected definition of the {\tt register} operation.

\begin{figure}
\begin{vdmsl}
    register : Name * [Email] ==> Id
    register(name, email) ==
        (dcl i:Id := NextId;
        NextId := NextId + 1;
        NameBook(i) := name;
        if email <> nil then EmailBook(i) := email;
        return i)
    post 
        NameBook = NameBook~ munion {RESULT |-> name}
        and (email = nil and EmailBook = EmailBook~
            or email <> nil and 
              EmailBook = EmailBook~ munion {RESULT |-> email});
\end{vdmsl}
\caption{Corrected specification of MemberBook}
\label{fig:memberbook-corrected}
\end{figure}
 
\subsection{Refactoring}
\begin{figure}
\begin{center}
\includegraphics[width=0.8\textwidth]{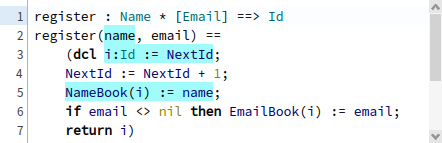}\\
{\bf (a) slice for NameBook}\\
\includegraphics[width=0.8\textwidth]{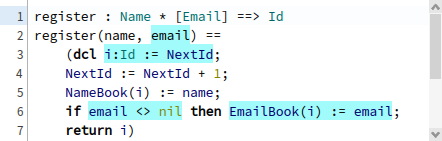}\\
{\bf (b) slice for EmailBook}\\
\includegraphics[width=0.8\textwidth]{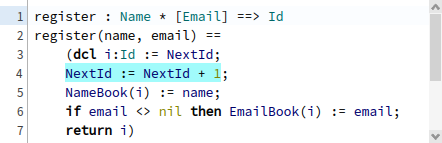}\\
{\bf (c) slice for NextId}\\
\end{center}
\caption{Slice for  the violated postcondition}
\label{fig:memberbook-refactoring-slices}
\end{figure}

The corrected {\tt register} operation shown in Fig. \ref{fig:memberbook-corrected} does two things: (1) to generate an id, and (2) to associate {\tt name} and optionally {\tt email} to the id.
We can look for the possibility of refactoring by splitting the operation.
Fig. \ref{fig:memberbook-refactoring-slices} shows slices for each state variable.
The slices do not overlap except Line 3, which is a simple renaming.
We can safely extract a {\tt generateId} operation to generate an id as shown in Fig. \ref{fig:memberbook-refactored}.

\begin{figure}
\begin{vdmsl}
generateId : () ==> Id
generateId() ==
    (dcl id:Id := NextId;
    NextId := NextId + 1;
    return id)
register : Name * [Email] ==> Id
register(name, email) ==
    let i = generateId()
    in
        (NameBook(i) := name;
        if email <> nil then EmailBook(i) := email;
        return i)
\end{vdmsl}
\caption{Refactored specification of MemberBook}
\label{fig:memberbook-refactored}
\end{figure}

\begin{figure}
\begin{center}
\includegraphics[width=0.8\textwidth]{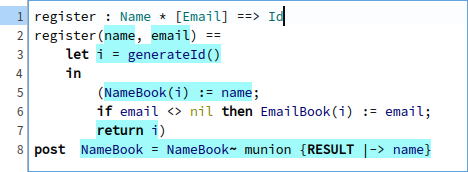}\\
\end{center}
\caption{Slice for  the loosen postcondition}
\label{fig:namebook-slice-loosen-post}
\end{figure}

Slicing can also be used to eliminate dead code.
The {\tt register} operation in Fig. \ref{fig:memberbook-refactored} registers the {\tt name} and optional {\tt email}.
Suppose that we modify the specification to register only {\tt name}.
We first loosen the postcondition of the {\tt register} operation to {\tt NameBook = NameBook\~{} munion \{RESULT $\mapsto$ name\}} by removing the other conjunct for {\tt EmailBook} and extract the slice for the new postcondition as shown in Fig. \ref{fig:namebook-slice-loosen-post}.
The {\tt if} statement in Line 6 is no longer necessary because it does not contribute to satisfying the postcondition.
Fig. \ref{fig:memberbook-eliminated} is the result of simplification.

\begin{figure}
\begin{vdmsl}
register : Name ==> Id
register(name) ==
    let i = generateId()
    in
        (NameBook(i) := name;
        return i)
post NameBook = NameBook~ munion {RESULT |-> name}
\end{vdmsl}
\caption{Simplified specification of MemberBook}
\label{fig:memberbook-eliminated}
\end{figure}

\section{Discussions}
\label{sec:discussions}

Specification slicing for VDM-SL is implemented in ViennaTalk~\cite{ViennaTalk}.
ViennaTalk is an IDE for VDM-SL specialized to the exploratory stage of the specification phase and is an open-source software published at github\footnote{\url{https://github.com/tomooda/ViennaTalk}}.
The implemented slicer is embedded in VDM Refactoring Browser~\cite{RefactoringBrowser} and the user can select a slicing criterion from the source and the extracted slice is shown highlighted in cyan.
Fig. \ref{fig:example-slice}, \ref{fig:memberbook-debug-slice}, \ref{fig:memberbook-refactoring-slices} and \ref{fig:namebook-slice-loosen-post} are screenshots of the VDM Refactoring Browser showing slices.
Based on the development and use of the slicer in ViennaTalk, we will discuss, in this section, specification slicing for VDM-SL from the perspectives of extraction algorithms and applications compared to program slicing.

Numerous practical programming languages allow aliasing; more than one reference to point at the same data.
Aliases make data dependency analysis complex because the extraction algorithms have to keep track of all possible aliases.
For example, the following python snippet prints {\tt 10} because the two variables {\tt l1} and {\tt l2} point at the same list object and the assignment to {\tt l2[0]} also overwrites {\tt l1[0]}, and therefore the slice for {\tt l1} must include {\tt l2[0]=10}.
\begin{lstlisting}{language=python}
l1 = [1,2,3]
l2 = l1
l2[0] = 10
print(l1[0])
\end{lstlisting}

On the other hand, VDM-SL has so-called value semantics that compound types, such as sets, maps, sequences and records, are values, not references, in assignments and pattern matchings.
The following block statement returns {\tt 1} although the statement looks similar to the Python snippet above.
The absence of aliases in VDM-SL makes slicing extraction simpler than slicing for programming languages with aliases.

\begin{lstlisting}{language=vdmsl}
(dcl l1 : seq of nat, l2 : seq of nat;
 l1 := [1,2,3]
 l2 := l1;
 l2(1) := 10;
 return l1(1))
\end{lstlisting}

Another significant difference from program slicing is assertions.
Although many programming languages provide assertions, they are often merely used as runtime sanity checking.
In VDM-SL, assertions such as invariants, preconditions and postconditions are theses of the specification; assertions declare properties that the system ought to have.
Assertions and their subexpressions are major concerns of the specification engineers and therefore they can be good sources of slicing criteria.
Fig. \ref{fig:memberbook-debug-slice} and Fig. \ref{fig:namebook-slice-loosen-post} are examples of such use of specification slicing, and slicing tools are expected to have UIs to specify an assertion or its subexpression as a slicing criterion.

\section{Concluding Remarks}
\label{sec:concluding-remarks}

Executable definitions of operations in VDM-SL have been used for testing and validation.
Specification slicing is another technique that takes benefit of executability by revealing potential dependencies among constituents.
While implicit definitions of operations also reveal influences via read and write access to state variables, specification slicing on explicit operations can reveal finer granularity of influences based on the semantics of statements.

We have investigated how the slicing technique used in imperative programming contributes to formal specifications, and are looking into further other applications of specification slicing.
For example, VDM-SL does not have specific features or standard tools to maintain the reusable library.
We expect slicing-based tools to support extracting necessary fragments of specification from libraries and migrating the retrieved fragments into the specification at hand.

\section*{Acknowledgements}
The authors thank valuable comments by anonymous reviewers.
A part of this research was supported by JSPS KAKENHI Grant Number JP 23K11058 and Nanzan University Pache Research Subsidy I-A-2 for the 2024 academic year.

\bibliographystyle{splncs03}

\bibliography{references}

\end{document}